\documentclass[12pt]{article}
\usepackage{amsmath}
\usepackage{amssymb}
\usepackage{amsfonts}
\usepackage{latexsym}
\usepackage{color}
\usepackage{graphicx}

\catcode `\@=11 \@addtoreset{equation}{section}
\def\theequation{\arabic{section}.\arabic{equation}}
\catcode `\@=12

\newcommand{\be}{\begin{equation}}
\newcommand{\en}{\end{equation}}
\newcommand{\bea}{\begin{eqnarray}}
\newcommand{\ena}{\end{eqnarray}}
\newcommand{\beano}{\begin{eqnarray*}}
\newcommand{\enano}{\end{eqnarray*}}
\newcommand{\bee}{\begin{enumerate}}
\newcommand{\ene}{\end{enumerate}}

\newcommand{\Hil}{{\cal H}}

\newcommand{\F}{{\cal F}}
\newcommand{\Lc}{{\cal L}}

\newcommand{\1}{1 \!\! 1}
\newtheorem{thm}{Theorem}

\newtheorem{lemma}[thm]{Lemma}

\newtheorem{defn}[thm]{Definition}

\begin{document}
\thispagestyle{empty}

\vspace*{1cm}

\begin{center}
{\Large \bf Quantum mechanical settings inspired by  RLC circuits}   \vspace{2cm}\\

{\large G. Alicata${}^1$, F. Bagarello${}^{1,2}$, F. Gargano${}^1$, S. Spagnolo${}^1$}\\
\vspace*{1cm}

\normalsize
${}^1$DEIM -Dipartimento di Energia, ingegneria dell' Informazione e modelli Matematici,
\\ Scuola Politecnica, Universit\`a di Palermo, I-90128  Palermo, Italy\\

\vspace*{.5cm}
${}^2$ INFN, Sezione di Napoli.\\

\end{center}

\vspace*{0.5cm}

\begin{abstract}
\noindent In some recent papers several authors used electronic circuits to construct loss and gain systems. This is particularly interesting in the context of PT-quantum mechanics, where this kind of effects appears quite naturally. The electronic circuits used so far are simple, but not so much. Surprisingly enough, a rather trivial RLC circuit can be analyzed  with the same perspective and it produces a variety of unexpected results, both from a mathematical and on a physical side. In this paper we show that this circuit produces two biorthogonal
bases associated to the Liouville matrix $\Lc$ used in the treatment of its dynamics, with a biorthogonality which is linked to the value of the parameters of the circuit. We also show that the related loss RLC circuit is naturally associated to a gain RLC circuit, and that the relation between the two is rather naturally encoded in $\Lc$. We propose a pseudo-fermionic analysis of the circuit, and we introduce the notion of $m$-equivalence between electronic circuits.
\end{abstract}

\vspace{2cm}


\vfill

\newpage

\section{Introduction}

$\mathcal{ PT}$-Quantum Mechanics, and its relatives, proved to be quite interesting both for the communities of mathematicians and of the physicists, \cite{ben1,ben,mosta,bagbook}. From a physical point of view, one of the most interesting applications of $\mathcal{ PT}$-symmetric systems have to do with gain-loss systems. One of the easiest way to analyse these systems is to construct electronic circuits which are easy to imagine and easy to manage. In fact, several authors have contributed in this area. Few references are \cite{circu1,circu2,circu3,circu4}. The circuits considered in these papers are not particularly complicated, by the point of view of their electronic content, but they naturally induce a dynamics that can be framed in the context of the quantum mechanics, giving rise to a quite rich framework. 

In what follows we will consider one of the easiest circuits one meets when studying electronics: the series RLC circuit. The differential equation for this circuit can be rewritten in a Schr\"odinger-like form, by introducing a suitable Hamiltonian $H$. The main aspect for us is that $H\neq H^\dagger$, and its eigenvalues are purely imaginary or complex, depending on the parameters of the circuit. 
This fact leads to the definition of the two  {\em phases} which are usually introduced in $\mathcal{ PT}$- Quantum Mechanics, that is
the \textit{unbroken} and \textit{broken} phase.
Then we use $H$ to construct biorthogonal bases of the Hilbert space $\Hil={\Bbb C}^2$, and we discuss what happens in the two situations, and we also analyse the possibility that coalescence of these eigenstates can arises in the \textit{exceptional points}.

We will also propose a pseudo-fermionic analysis of the mathematics of the circuit, based on the pseudo structures introduced in \cite{pf1}. In particular, we will show how $H$ can be factorized, and that this factorization produces a particular kind of ladder operators, at least in absence of exceptional points.
Our analysis suggests also to introduce the notion of $m$-equivalence between electronic circuits, and to analyse those circuits which are connected to different Hamiltonian-like operators connected by the adjoint operation. In particular, $H^\dagger$ produces a differential equation which can be interpreted as a {\em gain counterpart} of the original RLC circuit.

{The paper is organized as follows. In Sec.\ref{sect2}  we introduce the electronic circuit from which we deduce the quantum formalism and the phases of the system. The detailed analysis of the unbroken and broken phases, and of the exceptional points, is spread in Secs.\ref{sec::BP},\ref{sec::UP},\ref{sec::EP}, while the concept of $m$-equivalence is
proposed in Sec.\ref{anint}.
In Sec.\ref{PFs}  we  discuss a pseudo-fermionic framework for the Hamiltonian of the problem, and how the existence of pseudo-fermions is related to the (non)-appearance of exceptional points.
Finally, we conclude summarizing and discussing the main results exposed in the paper.}

\section{Stating the problem and first considerations}\label{sect2}

The starting point of our analysis is a simple $RLC$ circuit, with a resistor $R$, an inductor $L$ and a capacitor $C$, connected in series. There is no difference of potential applied in the circuit. Therefore, the equation of motion for the current $I(t)$ is just $\ddot I(t)+\frac{R}{L}\,\dot I(t)+\frac{I(t)}{LC}=0$ where $R,L,C$ are all assumed positive constants.
Equivalently, introducing $\alpha=\frac{R}{2L}$ and $\omega_0=\frac{1}{\sqrt{LC}}$, the equation for $I(t)$ can be rewritten as
\be
\ddot I(t)+2\alpha \dot I(t)+\omega_0^2 I(t)=0.
\label{21}\en
whose solution can be easily deduced, see Appendix. We will discuss now how, from (\ref{21}), it is possible to extract a number of results and considerations which are interesting in a quantum mechanical context. For that we first rewrite (\ref{21}) as the following, formal, Schr\"odinger-like equation:
\be
i\dot\Phi(t)=H\Phi(t), \qquad H=i\left(
                                   \begin{array}{cc}
                                     0 & 1 \\
                                     -\omega_0^2 & -2\alpha \\
                                   \end{array}
                                 \right), \quad \Phi(t)=\left(
                                                          \begin{array}{c}
                                                            x_1(t) \\
                                                            x_2(t) \\
                                                          \end{array}
                                                        \right),
                                                        \label{22}\en
where we have the constraint $x_2(t)=\dot x_1(t)$ and where we have identified $I(t)$ with $x_1(t)$: $x_1(t)=I(t)$. Then (\ref{22}) is not really a Schr\"odinger equation, but it still looks like a Schr\"odinger equation, and this is one of the reasons why we call it {\em formal}: in a {\em true} Schr\"odinger equation, $x_1(t)$ and $x_2(t)$ would not be related one to the other, of course.  What we have done here is just the standard trick which permits to write an $n$-th order differential equation as a system of $n$ first-order differential equations. We refer to the Appendix for some comments on the solutions of equations (\ref{21}) and (\ref{22}), and on their relation. We also refer to \cite{fernandez} for a similar approach for the damped harmonic oscillator. The second reason, and this is more interesting for us, is that $H$ is manifestly non-Hermitian: $H\neq H^\dagger$. It is interesting to notice that $H^\dagger$ can be {\em attached} to a circuit {\em reversing} the above procedure. More in details, if we consider the equation $i\dot \Psi(t)=H^\dagger\Psi(t)$, with $\Psi(t)=\left(
                                                  \begin{array}{c}
                                                    y_1(t) \\
                                                    y_2(t) \\
                                                  \end{array}
                                                \right)
$, after few computations we find the following differential equation for $y_1(t)$:
\be
\ddot y_1(t)-2\alpha \dot y_1(t)+\omega_0^2 y_1(t)=0,
\label{23}\en
which looks like the one in (\ref{21}) with $\alpha$ replaced by $-\alpha$ or, stated differently, with $R$ replaced by $-R$. This has an interesting interpretation since a negative resistor implies {\em gain}, whereas a positive resistor implies {\em loss}. Hence we can say that the $RLC$ circuit (associated to $H$) is a loss system, while the $-RLC$ circuit (corresponding to $H^\dagger$) is a gain system\footnote{The replacement $R\rightarrow -R$ is not the only way in which (\ref{23}) can be deduced from (\ref{21}). Another possibility is replace simultaneously $L$ and $C$ with $-L$ and $-C$. But we prefer to consider here the first, more {\em economical} choice. From a purely electronic point of view, replacing $R$ with $-R$ corresponds to replace a passive with an active element in the circuit.}.

It is worth noticing that, while (\ref{22}) is based on the assumption that $x_2(t)=\dot x_1(t)$, we are not requiring any a-priori relation between $y_1(t)$ and $y_2(t)$. Nevertheless, such a relation will appear because of the explicit expression of $H^\dagger$.

\vspace{2mm}

{\bf Remark:--} If we introduce the following linear combinations of $H$ and $H^\dagger$, $H_+=\frac{1}{2}(H+H^\dagger)$ and $H_-=\frac{1}{2}(H-H^\dagger)$, it is clear that $H_+=H_+^\dagger$ and $H_-=-H_-^\dagger$, and that $H=H_++H_-$. It is interesting to deduce the differential equations associated to these two operators. If we consider
$$
i\dot\eta^{(+)}(t)=H_+\eta^{(+)}(t), \qquad H_+=\frac{i}{2}\left(
\begin{array}{cc}
0 & 1+\omega_0^2 \\
\omega_0^2+1 & 0 \\
\end{array}
\right), \quad \eta^{(+)}(t)=\left(
\begin{array}{c}
\eta_1^{(+)}(t) \\
\eta_2^{(+)}(t) \\
\end{array}
\right),
$$
and
$$
i\dot\eta^{(-)}(t)=H_-\eta^{(-)}(t), \qquad H_+=\frac{i}{2}\left(
\begin{array}{cc}
0 & 1-\omega_0^2 \\
\omega_0^2-1 & -4\alpha \\
\end{array}
\right), \quad \eta^{(-)}(t)=\left(
\begin{array}{c}
\eta_1^{(-)}(t) \\
\eta_2^{(-)}(t) \\
\end{array}
\right),
$$
we find $$\ddot\eta_2^{(+)}(t)=-\frac{(1+\omega_0^2)^2}{4}\eta_2^{(+)}(t), \qquad\ddot\eta_2^{(-)}(t)+2\alpha \dot\eta_2^{(-)}(t)-\frac{(1-\omega_0^2)^2}{4}\eta_2^{(-)}(t)=0.$$
It is clear that $\eta^{(+)}(t)$ is a purely oscillating function, while $\eta^{(-)}(t)$ is not. This is in agreement with the fact that $H_+$ is Hermitian, while $H_-$ is anti-Hermitian. Hence $H_+$ can be thought to describe a circuit without any loss: an LC circuit. On the other hand, the presence of $R$ and -$R$ is all contained in $H_-$. Notice that the equation of motion for $\eta_2^{(-)}(t)$ simplifies when $\omega_0=1$, i.e. when $L=\frac{1}{C}$.

\vspace{2mm}

It is well known that different eigenstates of a non-Hermitian matrix $M$ are not orthogonal, in general, even if they belong to different eigenvalues. On the other hand, it is possible to check that the eigenstates of $M$ and those of $M^\dagger$ are biorthogonal. In fact, suppose that $Me_\alpha=\mu_\alpha e_\alpha$ and that $M^\dagger c_\alpha=\rho_\alpha c_\alpha$, for $\alpha\in I$. Here $I$ is a set which labels the  eigenstates of $M$ and $M^\dagger$. Now, since $\left<Me_\alpha,c_\beta\right>=\left<e_\alpha,M^\dagger c_\beta\right>$, and since $\left<Me_\alpha,c_\beta\right>=\overline{\mu_\alpha}\left<e_\alpha,c_\beta\right>$ and $\left<e_\alpha,M^\dagger c_\beta\right>=\rho_\beta\left<e_\alpha, c_\beta\right>$, we conclude that, for all $\alpha,\beta\in I$,
$$
\left<e_\alpha, c_\beta\right>(\overline{\mu_\alpha}-\rho_\beta)=0.
$$
Notice that we are considering here the possibility that the eigenvalues of $M$ and $M^\dagger$ are complex, since this is possible for our circuits $RLC$ and $-RLC$. The conclusion is straightforward: for all $\alpha$ and $\beta$ for which $\overline{\mu_\alpha}\neq\rho_\beta$ we must have $\left<e_\alpha, c_\beta\right>=0$: these two vectors are orthogonal. On the other hand, if $\overline{\mu_\alpha}=\rho_\beta$, $e_\alpha$ needs not being orthogonal to $c_\beta$.

Going back to our $H$ and $H^\dagger$, it is easy to find the following eigenstates:
\be
\varphi_\pm=N_{\varphi_\pm}\left(
                   \begin{array}{c}
                     1 \\
                     -i\lambda_\pm \\
                   \end{array}
                 \right), \qquad \Psi_\pm=N_{\Psi_\pm}\left(
                   \begin{array}{c}
                     1 \\
                     -i\frac{\mu_\pm}{\omega_0^2} \\
                   \end{array}
                 \right),
         \label{24}\en
where $N_{\varphi_\pm}$ and $N_{\Psi_\pm}$ are normalization constants to be fixed, while $\lambda_\pm$ and $\mu_\pm$ are the eigenvalues of $H$ and $H^\dagger$, respectively:
\be
H\varphi_\pm=\lambda_\pm\varphi_\pm, \qquad H^\dagger\Psi_\pm=\mu_\pm\Psi_\pm.
\label{25}\en
It is now convenient to consider two different situations: $\alpha>\omega_0$ and $\alpha<\omega_0$. With a slight abuse of language we will call the first case {\em the unbroken phase} ({\bf UP}) and the second {\em the broken phase} ({\bf BP}). The reason is that, as we will show in a moment, the eigenvalues of $iH$ and $(iH)^\dagger$ are real when  $\alpha>\omega_0$, while they are complex if  $\alpha<\omega_0$. This is somehow in the same direction of what is met in the literature on $\mathcal{ PT}$-quantum mechanics, \cite{ben,bagbook}. And in the same way we call {\em exceptional points} the eigenvalues (and the eigenvectors) arising when $\alpha=\omega_0$.

Then we have:

\begin{enumerate}

\item {\bf UP} ($\alpha>\omega_0$):
\be
\lambda_\pm=i\left(-\alpha\pm\sqrt{\alpha^2-\omega_0^2}\right),\qquad \mu_\pm=i\left(\alpha\pm\sqrt{\alpha^2-\omega_0^2}\right), \qquad \lambda_\pm=\overline{\mu_\mp}
\label{26}\en

\item {\bf BP} ($\alpha<\omega_0$):
\be
\lambda_\pm=-i\alpha\pm\sqrt{\omega_0^2-\alpha^2},\qquad \mu_\pm=i\alpha\pm\sqrt{\omega_0^2-\alpha^2}, \qquad \lambda_\pm=\overline{\mu_\pm}
\label{27}\en

\item {\bf EP} ($\alpha=\omega_0$)
\be
\lambda_{ep}:=\lambda_+=\lambda_-=-i\alpha,\qquad \mu_{ep}:=\mu_+=\mu_-=i\alpha, \qquad \lambda_{ep}=\overline{\mu_{ep}}
\label{28}\en

\end{enumerate}

Of course, in the {\bf EP}, the eigenvectors collapse and $H$ cannot be diagonalized. We will return on this particular case later on, and in particular in Section \ref{PFs}. For  the moment, we will not consider this case.  It is interesting to notice that biorthogonality of the sets $\F_\varphi=\{\varphi_\pm\}$ and $\F_\Psi=\{\Psi_\pm\}$ is different depending on the phase we are considering, and this is  in agreement with what we have discussed above for $M$ and $M^\dagger$. In particular, we have the following:

\vspace{2mm}

in the {\bf UP}, if we take $N_{\varphi_\pm}$ and $N_{\Psi_\pm}$ in such a way
\be
\overline{N_{\varphi_-}}\,N_{\Psi_+}=\left[1-\left(\frac{\sqrt{\alpha^2-\omega_0^2}+\alpha}{\omega_0}\right)^2\right]^{-1}, \quad \overline{N_{\varphi_+}}\,N_{\Psi_-}=\left[1-\left(\frac{\sqrt{\alpha^2-\omega_0^2}-\alpha}{\omega_0}\right)^2\right]^{-1},
\label{29}\en
then we have
\be
\left<\varphi_\pm,\Psi_\pm\right>=0, \qquad \left<\varphi_\pm,\Psi_\mp\right>=1
\label{210}\en
On the other hand, in the {\bf BP},  if we rather fix $N_{\varphi_\pm}$ and $N_{\Psi_\pm}$ in such a way
{ 
\be
\overline{N_{\varphi_-}}\,N_{\Psi_-}=\left[1+\left(\frac{\sqrt{\omega_0^2-\alpha^2}-i\alpha}{\omega_0}\right)^2\right]^{-1}, \quad \overline{N_{\varphi_+}}\,N_{\Psi_+}=\left[1+\left(\frac{\sqrt{\omega_0^2-\alpha^2}+i\alpha}{\omega_0}\right)^2\right]^{-1},
\label{211}\en
}
then we have
\be
\left<\varphi_\pm,\Psi_\pm\right>=1, \qquad \left<\varphi_\pm,\Psi_\mp\right>=0.
\label{212}\en
It is worth stressing that something of this kind was observed by one of us in a $\mathcal{ PT}$-version of the Graphene, \cite{baghat}, when discussing the eigenvectors of the Hamiltonian of that system. It is interesting for us to notice that similar behaviors also emerge in much simpler situations, like in our $RLC$ circuit.

Following now the standard approach discussed, for instance, in \cite{baginbagbook}, we could define two positive operators $S_\varphi$ and $S_\Psi$ as follows:
\be
S_\varphi f=\sum_{\alpha=\pm}\left<\varphi_\alpha,f\right>\varphi_\alpha, \qquad S_\Psi f=\sum_{\alpha=\pm}\left<\Psi_\alpha,f\right>\Psi_\alpha,
\label{212bis}\en
for all $f\in\Hil$, where $\Hil=\Bbb C^2$ endowed with its natural scalar product. These operators can be explicitly deduced and turn out to be
\be
S_\varphi=\left(
            \begin{array}{cc}
              |N_{\varphi_+}|^2+|N_{\varphi_-}|^2 & i\left(|N_{\varphi_+}|^2\overline{\lambda_+}+|N_{\varphi_-}|^2\overline{\lambda_-}\right) \\
              -i\left(|N_{\varphi_+}|^2{\lambda_+}+|N_{\varphi_+}|^2{\lambda_-}\right) & \left(|N_{\varphi_+}|^2|{\lambda_+}|^2+|N_{\varphi_-}|^2|{\lambda_-}|^2\right) \\
            \end{array}
          \right)
          \label{213}\en
and
{ 
\be
S_\Psi=\left(
            \begin{array}{cc}
              |N_{\Psi_+}|^2+|N_{\Psi_-}|^2 & i\left(|N_{\Psi_+}|^2\overline{\mu_+}+|N_{\Psi_-}|^2\overline{\mu_-}\right)/\omega_0^2 \\
              -i\left(|N_{\Psi_+}|^2{\mu_+}+|N_{\Psi_-}|^2{\mu_-}\right)/\omega_0^2 & \left(|N_{\Psi_+}|^2|{\mu_+}|^2+|N_{\Psi_-}|^2|{\mu_-}|^2\right)/\omega_0^4 \\
            \end{array}
          \right)
          \label{214}\en
          }
It is clear that these are both Hermitian matrices. It is less evident, but true, that they are also positive, see \cite{rs}. 

These two operators are one the inverse of the other, both in the  {\bf BP} and in the {\bf UP}. The reason is the following:

in the {\bf BP}, using (\ref{212}), we see that
\be
S_\varphi\Psi_\pm=\varphi_\pm, \qquad S_\Psi\varphi_\pm=\Psi_\pm.
\label{bp1}\en
In the {\bf UP}, using (\ref{210}), we rather find
\be
S_\varphi\Psi_\pm=\varphi_\mp, \qquad S_\Psi\varphi_\pm=\Psi_\mp.
\label{bp1bis}\en
Then, in both cases, 
$$
S_\Psi\,S_\varphi\Psi_\pm=\Psi_\pm, \qquad S_\varphi\,S_\Psi\varphi_\pm=\varphi_\pm,
$$
which imply that
\be
S_\varphi=S_\Psi^{-1}.
\label{bp3}\en
This is because both $\F_\varphi$ and $\F_\Psi$ are biorthogonal bases in $\Hil$: for all $f\in\Hil$ we have
\be
f=\sum_{\alpha=\pm}\left<\varphi_\alpha,f\right>\Psi_\alpha=\sum_{\alpha=\pm}\left<\Psi_\alpha,f\right>\varphi_\alpha,
\label{bp2}\en
in the {\bf BP}, while
\be
f=\sum_{\alpha=\pm}\left<\varphi_\alpha,f\right>\Psi_{\alpha_n}=\sum_{\alpha=\pm}\left<\Psi_\alpha,f\right>\varphi_{\alpha_n},
\label{bp2UP}\en
in the {\bf UP}. Here we use the following notation: if $\alpha=\pm$, then $\alpha_n=\mp$. Using the Dirac bra-ket notation, we rewrite (\ref{bp2}) and \eqref{bp2UP} respectively as the following resolutions of the identity:
$$
\1=\sum_{\alpha=\pm}|\varphi_\alpha\left>\right<\Psi_\alpha|=\sum_{\alpha=\pm}|\Psi_\alpha\left>\right<\varphi_\alpha|,
$$
in the {\bf BP}, and
$$
\1=\sum_{\alpha=\pm}|\varphi_{\alpha_n}\left>\right<\Psi_\alpha|=\sum_{\alpha=\pm}|\Psi_{\alpha_n}\left>
\right<\varphi_\alpha|,
$$
in the {\bf UP}.
Here $\1$ is the identity operator on $\Hil$, and $=(|f\rangle\langle g|)h=\left<g,h\right>f$, for all $f,g,h\in\Hil$.

\subsection{The {\bf BP}: $\alpha<\omega_0$}
\label{sec::BP}

In many quantum mechanical models it happens that $S_\varphi$ and $S_\Psi$ intertwine between the Hamiltonian of the system and its adjoint, see \cite{baginbagbook} and references therein. This is possible, since in those models these two operators are isospectral, i.e. they have the same eigenvalues. But this is not the case here, and then we do not expect that any such operator does exist. In fact, this can be explicitly proved: suppose that such a non zero operator $X$  exists: $HX=XH^\dagger$. Hence, taking the matrix elements of both sides of this equality in $\Psi_\beta$ and $\Psi_\alpha$ we should have $\left<\Psi_\beta, X\Psi_\alpha\right>(\overline{\mu_\beta}-\mu_\alpha)=0$, for all $\alpha,\beta=\pm$. But, see (\ref{27}), $\overline{\mu_\beta}\neq\mu_\alpha$ always. Therefore $\left<\Psi_\beta, X\Psi_\alpha\right>=0$ for all $\alpha,\beta$ and then, being $\F_\Psi$ a basis, $X=0$, which is against our assumption.

The fact that $H$ and $H^\dagger$ are not linked by any intertwining operator does not imply that there exist no other operator, $h$, related to $H$ and $H^\dagger$, for which this is possible. In particular, $h$ should have the same eigenvalues as those of $H$. Otherwise we fall into similar problems. Deducing such an operator $h$ is not particularly difficult. We first introduce the following anti-linear operator $U$:
\be
Uf=\sum_{\alpha=\pm}\left<f,\varphi_\alpha\right>\Psi_\alpha
\label{bp4}\en
 for all $f\in \Hil$. Due to (\ref{212}) it is clear that $U\Psi_\pm=\Psi_\pm$. However, since $U$ is antilinear, this does not imply that $U$ is the identity operator, \cite{herbut,bag2016}. In fact, for instance, $U(i\Psi_+)=-i\Psi_+$. Now, we can use $U$ to define the linear operator $h=UH^\dagger U$. Then
 \be
 h\Psi_\alpha=\lambda_\alpha\Psi_\alpha,
 \label{bp5}\en
for $\alpha=\pm$. Hence $h$ and $H$ are isospectral, but they have different eigenstates. It is now a simple exercise to prove that
 \be
HS_\varphi=S_\varphi h.
\label{bp6}\en
In fact, the two sides of this equation coincide on $\Psi_\alpha$: $HS_\varphi\Psi_\alpha=\lambda_\alpha\varphi_\alpha$ and $S_\varphi h\Psi_\alpha=\lambda_\alpha\varphi_\alpha$, $\alpha=\pm$. Then
\be
h=S_\Psi H S_\varphi,
\label{bp7}\en
and $h^\dagger=S_\varphi H^\dagger S_\Psi$. Formula \eqref{bp7} shows that $h$, originally introduced in terms of the antilinear operator $U$ as $UH^\dagger U$, can be rewritten without referring to $U$ at all as $S_\Psi H S_\varphi$ and, since $S_\varphi=S_\Psi^{-1}$, we conclude that $h$ and $H$ are similar. We deduce that
\be
S_\varphi H^\dagger=h^\dagger S_\varphi, \quad S_\Psi H=h S_\Psi, \quad H^\dagger S_\Psi=S_\Psi h^\dagger,
\label{bp6bis}\en
which are intertwining relations between $H$ and $h$, and between $H^\dagger$ and $h^\dagger$. The vectors in $\F_\varphi$ are eigenstates of $h^\dagger$, with eigenvalues $\mu_\alpha=\overline{\lambda_\alpha}$:
$$
h^\dagger\varphi_\alpha=S_\varphi H^\dagger S_\Psi\varphi_\alpha=S_\varphi H^\dagger \Psi_\alpha=\mu_\alpha S_\varphi \Psi_\alpha=
\overline{\lambda_\alpha}\varphi_\alpha.
$$
Here we have used formulas (\ref{25}) and (\ref{27}).

\subsubsection{An example}
In order to better understand the role of $h$ and $h^\dagger$, we fix the values of $\omega_0=1$ and $\alpha=\frac{1}{\sqrt{2}}$, choice which is compatible with the condition of being in the {\bf BP}. Hence we have
$$
H=i\left(
     \begin{array}{cc}
       0 & 1 \\
       -1 & -\sqrt{2} \\
     \end{array}
   \right), \qquad \lambda_\pm=\frac{-i\pm1}{\sqrt{2}}, \qquad \mu_\pm=\frac{i\pm1}{\sqrt{2}},
$$
with
$$
\varphi_\pm=N_{\varphi_\pm}\left(
                   \begin{array}{c}
                     1 \\
                     -i\lambda_\pm \\
                   \end{array}
                 \right), \qquad \Psi_\pm=N_{\Psi_\pm}\left(
                   \begin{array}{c}
                     1 \\
                     -i\mu_\pm \\
                   \end{array}
                 \right),
       $$
and $N_{\varphi_\pm}$ and $N_{\Psi_\pm}$ satisfying the following equalitites:
$$
\overline{N_{\varphi_+}}\,N_{\Psi_+}=\frac{1-i}{2}, \qquad \overline{N_{\varphi_-}}\,N_{\Psi_-}=\frac{1+i}{2}.
$$
If we now fix $N_{\Psi_+}=N_{\Psi_-}=1$, we easily deduce that
$$
S_\varphi=\left(
            \begin{array}{cc}
              1 & -1/\sqrt{2} \\
              -1/\sqrt{2} & 1 \\
            \end{array}
          \right), \qquad S_\Psi=\left(
            \begin{array}{cc}
              2 & \sqrt{2} \\
              \sqrt{2} & 2 \\
            \end{array}
          \right),
$$
which are one the inverse of the other, as expected. Then, using (\ref{bp7}) we get
$$
h=i\left(
     \begin{array}{cc}
       -\sqrt{2} & 1 \\
       -1 & 0 \\
     \end{array}
   \right).
$$
Notice that $h\neq h^\dagger$: in general, (\ref{bp7}) does not return an Hermitian operator.
It is now natural to ask which circuit this $h$ is attached to. For that, we write the Schr\"odinger equation $i\dot\eta(t)=h\eta(t)$, where $\eta(t)=\left(
                                                            \begin{array}{c}
                                                              w_1(t) \\
                                                              w_2(t) \\
                                                            \end{array}
                                                          \right)
$. Few computations show that this equation becomes $\ddot w_1(t)+\sqrt{2}\dot w_1(t)+w_1(t)=0$, which is exactly the same equation in (\ref{21}) for the same choice of parameters. It is also easy to check that the equation arising from $h^\dagger$ is nothing but the one in (\ref{23}). Hence we can conclude as follows: the pairs $(H,h)$ and $(H^\dagger,h^\dagger)$ produce the same differential equations for the current. The first pair describe a loss $RLC$ circuit, while the second pair describes a gain -$RLC$ circuit.

\subsection{An interlude: equivalence of ($RLC$-circuits)}\label{anint}

The previous example suggests that different Hamiltonians can be associated to the same equations of motion and, as a consequence, to different circuits which can be considered somehow related. For this reason, we introduce now the following definition:

\begin{defn}
Two circuits $C_a$ and $C_b$ are called $m$-equivalent if, when their equations of motion for the same variables are written in a Schr\"odinger-like form
$$
i\dot\Phi_a(t)=H_a\Phi_a(t),\qquad i\dot\Phi_b(t)=H_b\Phi_b(t),
$$
then $tr(H_a)=tr(H_b)$ and $\det(H_a)=\det(H_b)$.
\end{defn}

This definition is based on the following general result, which extend our conclusion in the example above: suppose that we the equation of the circuit can be written as $i\dot\Phi(t)=\hat H\Phi(t)$, for some $\hat H=\left(
                                                                                        \begin{array}{cc}
                                                                                          a & b \\
                                                                                          c & d \\
                                                                                        \end{array}
                                                                                      \right)
$ and $\Phi(t)=\left(
                 \begin{array}{c}
                   v_1(t) \\
                   v_2(t) \\
                 \end{array}
               \right)
$, with $v_2(t)=\dot v_1(t)$. Then it is straightforward to check that this corresponds to the following second order differential equation for $v_1(t)$:
\be
\ddot v_1(t)+i\,tr(\hat H)\dot v_1(t)-\det(\hat H)v_1(t)=0.
\label{AI1}\en
This implies that all the matrices with the same trace and determinant of $\hat H$ produce the same equation of motion.

\vspace{2mm}

{\bf Remark:--}
Being $m$-equivalent ($m$ stands for {\em mathematically}) does not mean necessarily that the two circuits are also {\em physically} equivalent, also because this latter notion should be defined, first. A deeper analysis of this aspect of our study is part of our future work.

\vspace{2mm}

It is well known that the trace and the determinant are invariant under similarity transformations. This is the case  when $H_a$ and $H_b$ are related by an invertible matrix $T$ via $H_b=TH_aT^{-1}$, which is exactly what happens in our example above. Hence, it is not surprising that $H$ and $h$ give rise to the same differential equation, as well as $H^\dagger$ and $h^\dagger$. However it may be useful to stress that matrices with the same traces and determinants need not being similar, as the counterexample with $A=\left(
                                                                   \begin{array}{cc}
                                                                     1 & 0 \\
                                                                     0 & 1 \\
                                                                   \end{array}
                                                                 \right)
$ and $B=\left(
                                                                   \begin{array}{cc}
                                                                     1 & 1 \\
                                                                     0 & 1 \\
                                                                   \end{array}
                                                                 \right)
$ clearly shows: $A$ and $B$ have the same traces (2) and the same determinant (1), but there exists no invertible matrix $R$ such that $B=RAR^{-1}$. However, we can easily check that $A$ and $B$ are related by an intertwining operator
$$
T=\left(
                                                                   \begin{array}{cc}
                                                                     a & b \\
                                                                     0 & 0 \\
                                                                   \end{array}
                                                                 \right),
$$
for arbitrary $a,b\in\Bbb R$: $BT=TA$. But, since $T^{-1}$ does not exist, we cannot conclude that $B=TAT^{-1}$. Hence a natural question arises: {\em suppose $H_b$ and $H_a$ are intertwined by some non invertible matrix $X$: $H_aX=XH_b$. Does these operators have necessarily the same traces and determinants?} Of course the answer would be affirmative if $\det(X)\neq0$, but when $\det(X)=0$ the answer is not. This can be understood by introducing the following counterexample:
$$
H_a=\left(
                                                                   \begin{array}{cc}
                                                                     2 & \alpha \\
                                                                     0 & \beta \\
                                                                   \end{array}
                                                                 \right),
                                                                 \quad H_b=\left(
                                                                   \begin{array}{cc}
                                                                     1 & 2 \\
                                                                     1 & 0 \\
                                                                   \end{array}
                                                                 \right), \quad X=\left(
                                                                   \begin{array}{cc}
                                                                     1 & 1 \\
                                                                     0 & 0 \\
                                                                   \end{array}
                                                                 \right).
$$
We have $H_aX=XH_b$, $\det(H_a)=2\beta$, $\det(H_b)=-2$, $tr(H_a)=2+\beta$, $tr(H_b)=1$. Hence we see that, independently of $\alpha$, traces and determinants of $H_a$ and $H_b$ coincide only if $\beta=-1$. Otherwise they are different.

\vspace{2mm}

The situation is much more complicated when we consider two coupled circuits, as in \cite{circu4} and references therein. Suppose our circuits are driven by the following coupled differential equations
\be
\left\{
    \begin{array}{ll}
\ddot x_1(t)+\alpha_1\,\dot x_1(t)+\alpha_2\,x_1(t)=\alpha_3\,x_2(t),\\
\ddot x_2(t)+\beta_1\,\dot x_2(t)+\beta_2\,x_2(t)=\beta_3\,x_1(t),\\
\end{array}
        \right.
\label{AI2}\en
where $x_1(t)$ and $x_2(t)$ are the dynamical variables (currents flowing in the two circuits, for instance) chosen to describe the system. This system can be written in a Schr\"odinger-like form by introducing the variables $y_j(t)=\dot x_j(t)$, $j=1,2$, and the following quantities:
$$
\Phi(t)=\left(
          \begin{array}{c}
            x_1(t) \\
            x_2(t) \\
            y_1(t) \\
            y_2(t) \\
          \end{array}
        \right), \qquad \Lc=\left(
                              \begin{array}{cccc}
                                0 & 0 & 1 & 0 \\
                                0 & 0 & 0 & 1 \\
                                -\alpha_2 & \alpha_3 & -\alpha_1 & 0 \\
                                \beta_3 & -\beta_2 & 0 & -\beta_1 \\
                              \end{array}
                            \right), \qquad i\dot\Phi(t)=H_{eff}\Phi(t),
$$
where $H_{eff}=i\Lc$.

From (\ref{AI2}) we can deduce the following fourth-order differential equation for $x_1(t)$ (and a similar computation could be repeated for $x_2(t)$):
\be
\ddddot x_1(t)-tr(\Lc) \dddot x_1(t)+\ddot x_1(t)(\alpha_2+\beta_2+\alpha_1\beta_1)+\dot x_1(t)(\alpha_1\beta_2+\alpha_2\beta_1)+\det(\Lc)x_1(t)=0.
\label{AI3}\en
Then  we have:
\begin{lemma}
Assume that $\alpha_2+\beta_2+\alpha_1\beta_1=\alpha_1\beta_2+\alpha_2\beta_1=0$. Then $\Lc$ and $S\Lc S^{-1}$ produce the same differential equation for $x_1(t)$, for all possible invertible matrix $S$.
\end{lemma}

Of course this Lemma could be stated in terms of $H_{eff}$ rather than $\Lc$. It is also clear that a similar result can be deduced for $x_2(t)$.


In \cite{circu3,circu4} the matrix $\Lc$ has essentially the following expression
$$
\Lc=\left(
                              \begin{array}{cccc}
                                0 & 0 & 1 & 0 \\
                                0 & 0 & 0 & 1 \\
                                -\alpha & \alpha\mu & \gamma & 0 \\
                                \alpha\mu & -\alpha & 0 & -\gamma \\
                              \end{array}
                            \right),
$$
which automatically satisfy $\alpha_1\beta_2+\alpha_2\beta_1=0$, while $\alpha_2+\beta_2+\alpha_1\beta_1=0$ only if $2\alpha=\gamma^2$. Hence we have concrete situations in which the hypothesis of the above Lemma are satisfied, and situations in which they are not.

Needless to say, the situation becomes harder and harder when we consider more coupled circuits. A detailed analysis of $m$-equivalence is part of our future projects.

\subsection{The {\bf UP}: $\alpha>\omega_0$}
\label{sec::UP}
The analysis of this regime is quite similar to that carried out for the {\bf BP}. Then, in this section, we will limit ourselves to highlighting some differences.
In the {\bf UP} case, we introduce the following new operators $T_\varphi$ and $T_\Psi$,  
\be
T_\varphi f=\left<\varphi_{+},f\right>\varphi_{-}+\left<\varphi_{-},f\right>\varphi_{+}, \qquad T_\psi f=\left<\psi_{+},f\right>\psi_{-}+\left<\psi_{-},f\right>\psi_{-},
\label{Top}\en
which work in the {\bf UP}, as a consequence of (\ref{210}), as $S_\varphi$ and $S_\Psi$ work in the {\bf BP}.
In particular, in the {\bf UP} we have 
\be
T_\varphi\Psi_\alpha=\varphi_\alpha, \qquad T_\Psi\varphi_\alpha=\Psi_\alpha,
\label{bp1UP}\en
for $\alpha=\pm$,
while in the {\bf BP} we have $T_\varphi\Psi_\pm=\varphi_\mp$ and $T_\Psi\varphi_\pm=\Psi_\mp$.

Hence, in both phases,
\be
T_\varphi=T_\Psi^{-1}.
\label{bp3UP}\en
Again, also in the {\bf UP} regime,  $T_\varphi$ and $T_\Psi$ cannot intertwine between $H$ and $H^\dagger$ since   these operators are not isospectral.
Nonetheless, using by the same procedure as before, we can deduce an isospectral operator $h$ also in {\bf UP}.
We first introduce the following anti-linear operator $\tilde U$:
\be
\tilde Uf=\left<f,\varphi_{+}\right>\psi_{-}+\left<f,\varphi_{-}\right>\psi_{+}
\label{bp4UP}\en
 for all $f\in \Hil$. Due to (\ref{210}) it is clear that $\tilde U\Psi_\pm=\Psi_\pm$. Then we define
 the linear operator $h=\tilde UH^\dagger \tilde U$, and we have 
 \be
 h\Psi_\alpha=\lambda_\alpha\Psi_\alpha,
 \label{bp5UP}\en
for $\alpha=\pm$. Hence, in this regime too, $h$ and $H$ are isospectral, but they have different eigenstates. Also, it is straightforward to prove that
 \be
HT_\varphi=T_\varphi h.
\label{bp6UP}\en
because the two sides of this equation coincide on $\Psi_+$ and $\Psi_-$: $HT_\varphi\Psi_\alpha=\lambda_\alpha\varphi_\alpha$ and $T_\varphi h\Psi_\alpha=\lambda_\alpha\varphi_\alpha$, $\alpha=\pm$. Then
\be
h=T_\Psi H T_\varphi,
\label{bp7UP}\en
and $h^\dagger=T_\varphi H^\dagger T_\Psi$. So, in complete analogy with the {\bf BP} regime, we have shown that $h$, originally introduced in terms of the antilinear operator $\tilde U$ as $\tilde UH^\dagger \tilde U$, can be rewritten   as $T_\Psi H T_\varphi$. We also find
\be
T_\varphi H^\dagger=h^\dagger T_\varphi, \quad T_\Psi H=h T_\Psi, \quad H^\dagger T_\Psi=T_\Psi h^\dagger,
\label{bp6bisUP}\en
which are intertwining relations between $H$ and $h$, and between $H^\dagger$ and $h^\dagger$. The vectors in $\F_\varphi$ are eigenstates of $h^\dagger$, with eigenvalues $\mu_\alpha=\overline{\lambda_\alpha}$:
$$
h^\dagger\varphi_\alpha=T_\varphi H^\dagger T_\Psi\varphi_\alpha=T_\varphi H^\dagger \Psi_\alpha=\mu_\alpha T_\varphi \Psi_\alpha=\mu_\alpha\varphi_\alpha.
$$
Here we have used formulas (\ref{25}) and (\ref{26}).

\subsection{The {\bf EP}: $\alpha=\omega_0$}
\label{sec::EP}

The \textbf{EP}  phase consists in the typical simultaneous coalescence of eigenvalues and  the corresponding eigenvectors of $H$.
This singular behavior, which has no counterpart in hermitian Hamiltonians, destroys the possibility of diagonalizing the Hamiltonian,
and it is accompanied by the so called \textit{self-orthogonality} condition, \cite{moise}.
In fact, for $\alpha=\omega_0$ the eigenvalues are given in \eqref{28}, and the eigenvectors are  
\be
\varphi_{EP}=N_{\varphi}\left(
                   \begin{array}{c}
                     1 \\
                     -\alpha \\
                   \end{array}
                 \right), \qquad \Psi_{EP}=N_{\Psi}\left(
                   \begin{array}{c}
                     1 \\
                     \alpha^{-1} \\
                   \end{array}
                 \right),
         \label{25bis}\en
from which the self orthogonality condition $\left<\varphi_{EP},\Psi_{EP}\right>=0$ follows.
Hence no intertwining can be defined between the two states, as we have done in the \textbf{UP} and \textbf{BP} phases. We will say more on \textbf{EP} in the next section.

\section{A pseudo-fermionic version of the circuit}\label{PFs}

In recent years a deformed version of the canonical anti-commutation relations (CAR) was proposed, leading to what were  called \textit{pseudo fermions}, \cite{pf1,FBgarg}.
The  functional structure is given in terms of biorthogonal bases $\F_\varphi=\{\varphi_-,\varphi_+\}$ and $\F_\Psi=\{\Psi_-,\Psi_+\}$, appearing together with lowering and raising operators defined by their action over $\F_\varphi$ and $\F_\varphi$. 
It was shown in \cite{FBgarg} that several two level systems, introduced in non hermitian quantum mechanics, can be easily represented in terms of  pseudo fermionic operators  which, therefore, provide a convenient general framework for those kind of systems.
The aim of this section is to give a full mathematical analysis of $H$ and $H^\dag$ in terms of pseudo fermionic operators.

Two operators $c,C$ on $\Hil={\Bbb C^2}$, with (in  general) $C\neq c^\dag$, are called pseudo fermions
if thye satisfy the following anti-commutation rules:
\begin{eqnarray}
\{c,C \}=\1, \quad \{c,c \}=\{C,C \}=0,
\label{eCAR}
\end{eqnarray}
where $\{A,B\}=AB+BA$ is the anticommutator of two operators $A,B$.
It was shown in \cite{pf1} that the general expressions of $c,C$ are the following:
\bea
c=a_{12}\left(
               \begin{array}{cc}
                 a & 1 \\
                 -a^2 & -a \\
               \end{array}
             \right), \qquad C=b_{12}\left(
               \begin{array}{cc}
                 b & 1 \\
                 -b^2 & -b \\
               \end{array}
             \right),\label{pfermions}
\ena
where $a,b,a_{12},b_{12}\in \mathbb{C}$ and $(a-b)\gamma=1$ with $\gamma=a_{12}b_{12}(b-a)$.
The condition $(a-b)\gamma=1$ is the existence condition for the pseudo fermionic operators $c,C$: when this equality is not satisfied, no pair of operators satisfying (\ref{eCAR}) can be introduced.

A rather general diagonalizable Hamiltonian that can be written out of $c,C$ is
\be
H_{PF}=\omega Cc+\rho \1=\left(
               \begin{array}{cc}
                 \omega\gamma a+\rho & \omega\gamma \\
                 -\omega\gamma ab & -\omega\gamma b+\rho \\
               \end{array}
             \right),\label{Hpf}
\en
where $\omega,\rho\in \mathbb{C}$.
We now show that the Hamiltonian in \eqref{22} can be written as in \eqref{Hpf}, with an appropriate choice of the parameters of $H_{PF}$. The choice also depends on the phase of the system (\textbf{UP}, \textbf{BP}).
We immediately stress here that no pseudo fermionic representation can be given in the \textbf{EP} phase.
In fact, as shown in the following, in presence of \textbf{EP} the existence condition $(a-b)\gamma=1$ cannot be satisfied anymore.

\subsection{The \textbf{UP} case $\alpha>\omega_0$}\label{sectPF1}

In this case we can write the Hamiltonian $H$ in \eqref{22} as $H_{PF}$ with the following identifications:
 \be\label{r1}
 a_{\pm}=\alpha\pm \sqrt{\alpha^2-\omega_0^2},\quad 
  b_{\pm}=\alpha\mp \sqrt{\alpha^2-\omega_0^2},\quad
   \rho_{\pm}=-i\alpha\mp i \sqrt{\alpha^2-\omega_0^2},\quad
   \omega_{\pm}\gamma_{\pm}=i.
 \en
 The subscript $\pm$  indicates that we have  two possible choices of the parameters.
 Moreover, to ensure that the pseudo fermionic existence condition $(a_{\pm}-b_{\pm})\gamma_\pm=1$ is satisfied, we need to require that
 $a_{\pm}\neq b_{\pm}$, which can be satisfied only if $\alpha\neq\omega_0$: \textbf{EP} are not allowed, then. 
 {In the latter case we obtain
 that $\gamma_{\pm}=(a_{\pm}-b_{\pm})^{-1}=\left(\pm2\sqrt{\alpha^2-\omega_0^2}\right)^{-1}$ and,  from the last condition in \eqref{r1}, $\omega_{\pm}=\pm2i\left(\sqrt{\alpha^2-\omega_0^2}\right)$.}
 
To simplify our mathematical treatment, we observe that $a_\pm=b_\mp$. Hence, when going from the $'+'$ to the $'-'$ choice in (\ref{r1}), $c$ becomes $C$,  the eigenvalues $\rho_{+},(\omega_{+}+\rho_{+})$ become $(\omega_{-}+\rho_{-}),\rho_{-}$, and the eigenstates $\varphi_{+},\Psi_+$ are replaced by $\varphi_{-},\Psi_-$ (viceversa if we pass from $'-'$ to  $'+'$). Hence the Hamiltonian $H_{PF}$ has essentially the same structure for both choices of signs. Therefore, hereafter, we shall omit the subscripts in the quantity \eqref{r1}  and focus only to the $'+'$ case.

The eigensystems  of $H_{PF},H_{PF}^\dag$,
$$
H_{PF}\varphi_{\pm}=\lambda_{\pm}\varphi_{\pm}, \quad H_{PF}^\dag\Psi_{\pm}={\mu}_{\pm}\Psi_{\pm},
$$
are the following:
 $\lambda_-=\rho$ and $\lambda_+=\omega+\rho$, $\mu_\pm=\overline{\lambda_{\mp}}$, while the
 eigenstates $\varphi_{\pm},\Psi_{\pm}$ are those in \eqref{24}, with the normalization conditions given in \eqref{29} which now read as:
$$
\overline{N_{\varphi_-}}\,N_{\Psi_+}=\left(1-\frac{a^2}{\omega_0^2}\right)^{-1},\quad 
\overline{N_{\varphi_+}}\,N_{\Psi_-}=\left(1-\frac{b^2}{\omega_0^2}\right)^{-1}.
$$

It is easy to show that the pseudofermionic operators $c,C$ can be easily written in a bra-ket form as 
 $$c=|\varphi_-\rangle\langle\Psi_-|,\quad C=|\varphi_+\rangle\langle\Psi_+|,$$
 
 where  $a_{12}=\frac{\overline{N_{\Psi_-}}\,N_{\varphi_-}}{a},\quad b_{12}=\frac{\overline{N_{\Psi_+}}\,N_{\varphi_+}}{b}$ in \eqref{pfermions}. 
 
 Hence, as already stated, $c,C$ act like lowering and raising operators on {$\F_{\varphi}$ and $\F_{\Psi}$}, since they satisfy 
 \bea
 c\varphi_-&=&0,\quad c\varphi_+=\varphi_-,\quad C\varphi_-=\varphi_+,\quad C\varphi_+=0,\label{cc1}\\
 C^\dag \Psi_-&=&0,\quad C^\dag\Psi_+=\Psi_-,\quad c^\dag \Psi_-=\Psi_+,\quad c^\dag\Psi_+=0.\label{cc2}
 \ena
 Moreover, introducing $\mathcal{N}_{\varphi}=Cc$ and $\mathcal{N}_{\Psi}=c^\dag C^\dag$, these behave like non-Hermitian number operators:
\bea
\mathcal{N}_{\varphi}\,\varphi_-&=&0,\quad \mathcal{N}_{\varphi}\,\varphi_+=\varphi_+,\label{numb1}\\
  \mathcal{N}_{\Psi}\Psi_-&=&0,\quad  \mathcal{N}_{\Psi}\Psi_+=\Psi_+\label{numb2}.
\ena

In \cite{baginbagbook} it is discussed in details how pseudo-fermions can be connected, via some suitable similarity relation, to ordinary fermions, i.e. to operators $A$ and $A^\dagger$ satisfying CAR: $\{A,A^\dagger\}=\1$, with $A^2=0$. It is easy to check that the same similarity condition relates $H$ (which here coincides with $H_{PF}$) to the Hamiltonian of a fermionic harmonic oscillator, $H_{fho}=\omega A^\dag A+\rho\1$. First of all we have $A=S_{\Psi}^{1/2}cS_{\varphi}^{1/2}$ and $A^\dag=S_{\Psi}^{1/2}C S_{\varphi}^{1/2}$. Notice that these positive square roots exist, and are unique, since both $S_\varphi$ and $S_\Psi$ are Hermitian and positive, \cite{rs}. Furthermore, if we introduce the vectors $e_\pm=S_{\Psi}^{1/2}\varphi_\pm=S_{\varphi}^{1/2}\Psi_\pm$, the biorthonormal conditions \eqref{210} imply that $\langle e_-,e_- \rangle=\langle e_+,e_+ \rangle=1$ and $\langle e_-,e_+ \rangle=0$. Then $\F_{e}=\{e_-,e_+\}$ is a orthonormal basis of $\Hil$.

Using \eqref{cc1},\eqref{cc2}, and the formulas for $A$ and $A^\dagger$ we easily see that
$$
Ae_-=0,\qquad
Ae_+=e_-,\qquad
A^\dag e_-=e_+,\qquad
A^\dag e_+=0,
$$
which show that $A,A^\dag$ are  lowering and raising operators for $\F_e$. 
Of course $A^2=(A^\dagger)^2=0$. Moreover
$$
\{A,A^\dagger\}=S_{\Psi}^{1/2}\{c,C\}S_{\varphi}^{1/2}=\1,
$$

A straightforward consequence of these result is that the similarity between $A$, $c$ and $C$ extends to $H$ and $H_{fho}$:
$$
H=\omega Cc+\rho\1=S_{\Psi}^{1/2}\left(\omega A^\dag A+\rho \1\right) S_{\varphi}^{1/2}=S_{\Psi}^{1/2}H_{fho}S_{\varphi}^{1/2}
$$

Our analysis in Section \ref{anint} shows that $H_{fho}$ is linked to the same RLC circuit as $H$. This is not surprising since, even if $H_{fho}$ looks formally Hermitian (and, in this perspective, it is strange that it describes a loss circuit), in fact it is not, since $\omega$ and $\rho$ are both complex.

\vspace{2mm}

{\bf Remark:--} The pseudo-fermionic approach proposed here is interesting also because it shows that $H$ can be (essentially) factorized in terms of ladder operators. And when an Hamiltonian can be factorized, one can look at its SUSY partner, $H^S$. In this case we have $H^S=\omega c C+\rho\1$, and we can easily check the following: while $H\varphi_{-}=\rho\varphi_{-}$ and $H\varphi_{+}=(\rho+\omega)\varphi_{+}$, for $H^S$ the role of the eigenvalues is inverted:
$$
H^S\varphi_{-}=(\rho+\omega)\varphi_{-}, \qquad H^S\varphi_{+}=\rho\varphi_{+}.
$$


 \subsection{The \textbf{BP} case $\alpha<\omega_0$}
 
 In this case to identify the Hamiltonian $H$ with $H_{PF}$, it is enough to consider the following choice of the parameters:
  \be\label{r11}
  a_{\pm}=\alpha\pm i\sqrt{\omega_0^2-\alpha^2},\quad
   b_{\pm}=\alpha\mp i\sqrt{\omega_0^2-\alpha^2},\quad
    \rho_{\pm}=-i\alpha\pm  \sqrt{\omega_0^2-\alpha^2},\quad
    \omega_{\pm}\gamma_{\pm}=i.
  \en
 The pseudo fermionic existence condition $(a_{\pm}-b_{\pm})\gamma_\pm=1$ is also in satisfied in the \textbf{BP} case, as $a_{\pm}\neq b_{\pm}$ is surely true whenever  $\alpha<\omega_0$. Hence
   $\gamma_{\pm}=(a_{\pm}-b_{\pm})^{-1}=\left(\pm2i\sqrt{\omega_0^2-\alpha^2}\right)^{-1}$, and  from \eqref{r1} $\omega_{\pm}=\mp2\left(\sqrt{\omega_0^2-\alpha^2}\right)$.

 The eigensystems of $H_{PF},H_{PF}^\dag$ now satisfy \eqref{212}, and  $c,C$ can be easily written in terms of $\F_\varphi$ and $\F_\Psi$ as 
 $$c=|\varphi_-\rangle\langle\Psi_+|,\quad C=|\varphi_+\rangle\langle\Psi_-|,$$
which is different from their expression in the {\bf UP}. As in the \textbf{UP} case, $c,C$ and their related number operators  $\mathcal{N}_{\varphi}=Cc,\mathcal{N}_{\Psi}=c^\dag C^\dag$
behaves like in \eqref{cc1}-\eqref{cc2} and \eqref{numb1}-\eqref{numb2}. In fact, no essential difference exists between the two phases, and all what deduced in Section \ref{sectPF1} can be repeated here, with only minor changes.

\subsubsection{A remark on the $\mathcal{P}\mathcal{T}$-symmetry}

In the literature on $\mathcal{P}\mathcal{T}$-Quantum Mechanics,
 the parity and the time reversal operators $\mathcal{P}$ and $\mathcal{T}$ are usually defined by
\bea
\mathcal{P}=\left(
\begin{array}{cc}
	0 & 1 \\
	1 & 0 \\
\end{array}
\right) \quad \mathcal{T} := \text{complex conjugate}.\label{PT}
\ena
With this choice, the Hamiltonian $H$ in (\ref{22}) is, in general, not  $\mathcal{ PT}$-symmetric because $[\mathcal{P} \mathcal{T},H]\neq0$. In fact, for instance,
$$
H\mathcal{PT}\left(
\begin{array}{cc}
 1 \\
 0 \\
\end{array}
\right)=i\left(
\begin{array}{cc}
1 \\
-2\alpha \\
\end{array}
\right), \qquad \mathcal{PT}H\left(
\begin{array}{cc}
1 \\
0 \\
\end{array}
\right)=i\left(
\begin{array}{cc}
\omega_0^2 \\
0 \\
\end{array}
\right).
$$
A similar result can be deduced acting on $\left(
\begin{array}{cc}
0 \\
1 \\
\end{array}
\right)$. Hence, a necessary and sufficient condition for $H$ to be $\mathcal{ PT}$-symmetric is that $\omega_0=1$ and that $\alpha=0$.  Then, our RLC circuit cannot produce a $\mathcal{ PT}$-symmetric Hamiltonian except when there is no resistance at all. In particular,  \eqref{21} becomes the equation of an classical harmonic oscillator and our current oscillates, and shows no damping.

The same result can be deduced, of course, working with $H_{PF}$. 
In order for $H_{PF}$ to be  
$\mathcal{P}\mathcal{T}$-symmetric, we require that $[\mathcal{P} \mathcal{T},H_{PF}]=0$, a condition which is satisfied again under the same conditions on $\alpha$ and $\omega_0$, which can only occur in the \textbf{BP} phase.
In this case  \eqref{r11} reduces
to $a_{\pm}=\pm i, b=\mp i, \rho=\pm1, \omega_{\pm}\gamma_{\pm}=1$ with $\gamma_{\pm}=\mp i/2,\omega=\mp2$,
from which in particular we deduce that all the eigenvalues are real.

\section{Conclusions}

In this paper we have shown how a simple RLC circuit gives rise to several interesting mathematical results, on the line of $\mathcal{ PT}$-Quantum Mechanics, when the equation of the current of the circuit is written in a Schr\"odinger-like form, by means of a suitable, manifestly non-Hermitian Hamiltonian $H$. Our analysis extends significantly what was discussed in \cite{fernandez}, in connection with a damped oscillator. In particular we have seen that $H^\dagger$ produce a gain circuit, which differs from the original circuit since $R$ is replaced by $-R$. Biorthogonal bases have been constructed, but the explicit form of the biorthogonality condition is linked to the particular phase of the system, broken or unbroken. We have also proposed a pseudo-fermionic analysis of our circuit, analysis which allows for an interesting factorization of the Hamiltonian of the circuit.

Motivated by our analysis, we have introduced the definition of $m$-equivalent circuits for our particularly simple system, and we have discussed how this notion can be extended to more general circuits. This particular aspect of our analysis will be considered further in our future work.

We should finally observe that no substantial difference appears if the series RLC circuit is replaced by a parallel RLC circuit, since the duality relationship of electrical circuits produces again a second order differential equation similar to (\ref{21}).

\renewcommand{\theequation}{A.\arabic{equation}}

\section*{Appendix:  Solution of the equation (\ref{21}) and (\ref{22})}

As we have discussed in Section \ref{sect2}, equation (\ref{22}) is just a different way to write (\ref{21}). And in fact, as we will briefly discuss here, both give the same solution for the current $I(t)$. For concreteness, we will only consider the case $\alpha<\omega_0$, i.e. the {\bf BP}. The solution of (\ref{21}) can be deduced in a very standard way. Assuming the following initial conditions,
$$
I(0)=I_0, \qquad \dot I(0)=-\alpha I_0-V_0/L,
$$
where $V_0=\frac{1}{C}\int_{-\infty}^0I(t)\,dt$, we find
\be
I(t)=e^{-\alpha t}\left(I_0\cos(\omega_dt)-\frac{V_0}{L\omega_d}\sin(\omega_dt)\right).
\label{a1}\en
On the other hand, equation (\ref{22}) admits the following solution: $\Phi(t)=e^{-iHt}\Phi(0)$. But, since $\varphi_{\pm}$ are eigenvectors of $H$, and since $\F_\varphi$ is a basis for $\Hil$, it follows that $\Phi(0)=b_+\varphi_++b_-\varphi_-$, where $b_{\pm}$ will be fixed by the initial conditions on $I(t)$ and $\dot I(t)$. Hence
$$
\Phi(t)=e^{-iHt}\left(b_+\varphi_++b_-\varphi_-\right)=b_+e^{-i\lambda_+t}\varphi_++b_-e^{-i\lambda_-t}\varphi_-=\left(
                                                          \begin{array}{c}
                                                            x_1(t) \\
                                                            x_2(t) \\
                                                          \end{array}
                                                        \right).
$$
Hence, since $x_1(t)$ in our procedure is exactly the current, see Section \ref{sect2}, we only need to compute the first component of $\Phi(t)$. Recalling (\ref{24}), it is easy to check that this function $x_1(t)$ coincides, as expected, with the function $I(t)$ given in (\ref{a1}).


\end{document}